\documentclass[prl,showpacs,twocolumn,preprintnumbers,amsmath,groupedaddress,amssymb,superscriptaddress,floatfix]{revtex4}
\usepackage[dvips]{graphicx}
\usepackage[latin1]{inputenc}
\usepackage{xcolor,bm}
\usepackage{nicefrac}

\begin{document}
\draft

\hyphenation{a-long}

\title{Influence of hydrostatic pressure on the bulk magnetic properties of Eu$_{2}$Ir$_{2}$O$_{7}$}

\author{G.~Prando}\email[E-mail: ]{g.prando@ifw-dresden.de}\affiliation{Center for Transport and Devices of Emergent Materials, Technische Universit\"at Dresden, D-01062 Dresden, Germany}\affiliation{Leibniz-Institut f\"ur Festk\"orper- und Werkstoffforschung (IFW) Dresden, D-01171 Dresden, Germany}
\author{R.~Dally}\affiliation{Department of Physics, Boston College, Chestnut Hill, Massachusetts 02467, USA}
\author{W.~Schottenhamel}\affiliation{Leibniz-Institut f\"ur Festk\"orper- und Werkstoffforschung (IFW) Dresden, D-01171 Dresden, Germany}
\author{Z.~Guguchia}\affiliation{Laboratory for Muon Spin Spectroscopy, Paul Scherrer Institut, CH-5232 Villigen PSI, Switzerland}
\author{S.-H.~Baek}\affiliation{Leibniz-Institut f\"ur Festk\"orper- und Werkstoffforschung (IFW) Dresden, D-01171 Dresden, Germany}
\author{R.~Aeschlimann}\affiliation{Leibniz-Institut f\"ur Festk\"orper- und Werkstoffforschung (IFW) Dresden, D-01171 Dresden, Germany}
\author{A.~U.~B.~Wolter}\affiliation{Leibniz-Institut f\"ur Festk\"orper- und Werkstoffforschung (IFW) Dresden, D-01171 Dresden, Germany}
\author{S.~D.~Wilson}\affiliation{Department of Materials, University of California, Santa Barbara, California 93106, USA}
\author{B.~B\"uchner}\affiliation{Leibniz-Institut f\"ur Festk\"orper- und Werkstoffforschung (IFW) Dresden, D-01171 Dresden, Germany}\affiliation{Institut f\"ur Festk\"orperphysik, Technische Universit\"at Dresden, D-01062 Dresden, Germany}
\author{M.~J.~Graf}\affiliation{Department of Physics, Boston College, Chestnut Hill, Massachusetts 02467, USA}

\widetext

\begin{abstract}
We report on the magnetic properties of Eu$_{2}$Ir$_{2}$O$_{7}$ upon the application of hydrostatic pressure $P$ by means of macroscopic and local-probe techniques. In contrast to previously reported resistivity measurements, our dc magnetization data unambiguously demonstrate a non-monotonic $P$-dependence for $T_{N}$, i.~e., the critical transition temperature to the magnetic phase. More strikingly, we closely reproduce the recently calculated behaviour for $T_{N}$ under the assumption that $P$ lowers the $U/W$ ratio (i.~e., Coulomb repulsion energy over electronic bandwidth). Zero-field muon-spin spectroscopy measurements confirm that the local magnetic configuration is only weakly perturbed by low $P$ values, in agreement with theoretical predictions. The current results experimentally support the preservation of a $4$-in/$4$-out ground state and, simultaneously, a departure from the single-band $j_{\text{eff}} = 1/2$ model across the accessed region of the phase diagram.
\end{abstract}

\pacs{62.50.-p, 71.30.+h, 75.40.Cx, 76.75.+i}

\date{\today}

\maketitle

\narrowtext

The arrangement of magnetic moments at the vertices of a pyrochlore lattice (i.~e., a geometrically-frustrating crystalline structure composed of corner-sharing tetrahedra) leads to a great variety of exotic electronic ground states for $R_{2}M_{2}$O$_{7}$ materials \cite{Gar10}. One important finding common to several families of these oxides is that changes in $r_{I}$, the ionic radius of the rare-earth ion $R^{3+}$, gradually tune the local crystalline environment around the transition metal ion $M^{4+}$ and the overall electronic behaviour of the compound, accordingly. This is verified for $M$ ions belonging to the $4d$ (e.~g., Mo) \cite{Kez06,Pra14} and $5d$ (e.~g., Ir) \cite{Yan01,Mat11,Wit14,Tak14} groups. Here, the characteristic temperature $T_{\text{MI}}$ for the development of a MIT (metal-to-insulator transition) is directly controlled by the average value $\langle r_{I}\rangle$ related to a gradual $R_{1-x}R_{x}^{\prime}$ substitution \cite{Kat00,Mor01,Han07,Ued15}. However, a chemical dilution is generally delicate to handle as it necessarily introduces a non-negligible degree of quenched disorder, possibly detrimental for most of the intrinsic physical properties under investigation. Moreover, $R$ ions may also be characterized by localized magnetic moments arising from their incomplete $f$ electronic shells and this fact may influence measurements when the actual magnetic properties are being studied.

These aspects are particularly relevant as the MIT in $R_{2}$Mo$_{2}$O$_{7}$ and $R_{2}$Ir$_{2}$O$_{7}$ is associated with a dramatic change in the magnetic behaviour as well. Antiferromagnetic insulating and metallic paramagnetic states compose indeed the two main regions of the EPD (electronic phase diagram) of Ir-based pyrochlores \cite{Mat11}, even if this is only a very simplistic description. Theory predicts that comparable energy scales for the Coulomb repulsion $U$ and the spin-orbit coupling make the EPD of pyrochlore iridates extremely rich with exotic states. In the insulating regime, a non-collinear state is theoretically predicted \cite{Wan11,Wit12,Wit13} and experimentally claimed \cite{Sag13,Dis14} for Ir magnetic moments, pointing along the local $\langle 111 \rangle$ directions all inside or all outside the tetrahedron at whose vertices they sit ($4$-in/$4$-out configuration). Theory suggests that such an arrangement is necessary for the achievement of topologically non-trivial properties such as, e.~g., the Weyl semimetallic phase \cite{Wan11,Sus15}.

A cleaner structural approach in order to fine-tune the ground state of the system is to vary $P$ (external pressure). This method has been very successful for $M$ = Mo \cite{Ish04,Ape06,Mir06,Igu09} and, more recently, for the case of Ir \cite{Sak11,Taf12} where $P$ gradually suppresses the insulating state and triggers metallicity in several compounds \cite{Taf12,Ued15}. Transport properties have been mainly discussed in the literature and, although the trend is clear, a purely metallic phase is never properly achieved. Moreover, discrepancies exist among the results for different compounds. While $T_{\text{MI}}$ is clearly sensitive to pressure for Sm$_{2}$Ir$_{2}$O$_{7}$, Nd$_{2}$Ir$_{2}$O$_{7}$, and Pr$_{2}$Ir$_{2}$O$_{7}$ (and intermediate compositions) \cite{Ued15}, no sizeable effect of $P$ is reported for this quantity in Eu$_{2}$Ir$_{2}$O$_{7}$ \cite{Taf12}. Finally, a detailed investigation of the magnetic properties of these compounds under $P$ is still missing.
\begin{figure*}[htbp]
	\vspace{5.4cm} \includegraphics{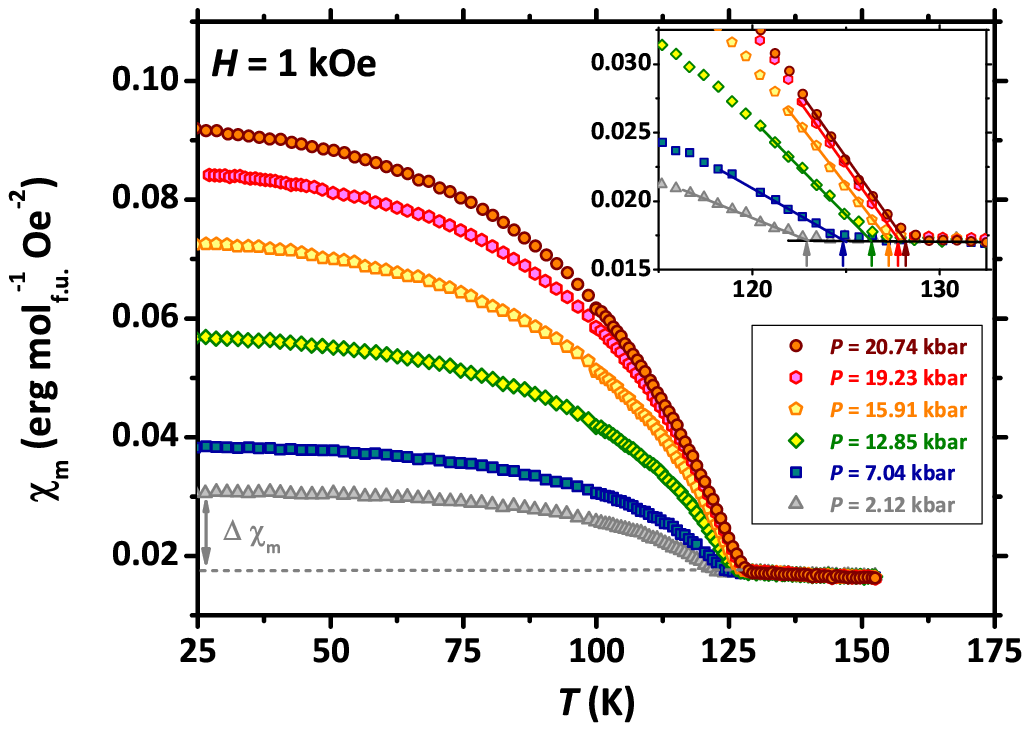} \includegraphics{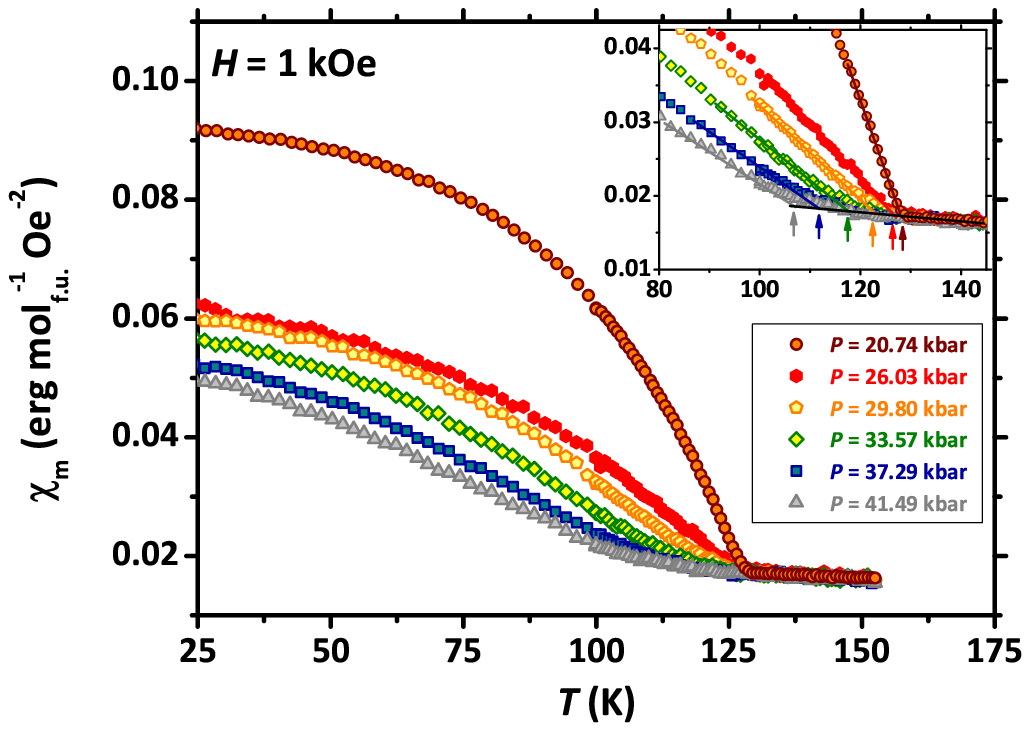}
	\caption{\label{GraGaskets}(Color online) Main panels: $T$ dependence of the FC susceptibility $\chi_{m}$ for $P \lesssim 20$ kbar and $P \gtrsim 20$ kbar after background subtraction (left and right panel, respectively). The amount of FC susceptibility at $25$ K related to the magnetic phase of Eu$_{2}$Ir$_{2}$O$_{7}$ is defined as $\Delta\chi_{m}$ (see the graphical definition in the left panel). The insets display enlargements of the magnetic transition region together with the double-linear fits defining the actual $T_{N}$ values (arrows).}
\end{figure*}

In this paper we focus on the magnetic properties of Eu$_{2}$Ir$_{2}$O$_{7}$ under pressure, both from macroscopic (dc magnetometry) and local ($\mu^{+}$SR, muon-spin spectroscopy) perspectives. The absence of a localized magnetic moment from $f$ shells in Eu$_{2}$Ir$_{2}$O$_{7}$ is a great advantage in the study of the intrinsic magnetic properties of the Ir sublattice. From our magnetization measurements, we deduce a clearly non-monotonic $P$-dependence of $T_{N}$ pointing to its departure from $T_{\text{MI}}$ \cite{Taf12}. As striking result, we closely reproduce the behaviour recently reported for $T_{N}$ from relativistic LDA (local density approximation) + DMFT (dynamical mean-field theory) calculations \cite{Shi15} under the assumption that $P$ linearly influences the $U/W$ ratio ($W$ representing the electronic bandwidth). By confirming this theoretical scenario, our measurements support the departure from the single-band $j_{\text{eff}} = 1/2$ model for iridates \cite{Wit14,Kim08}, in favour of a hybridized state between $j_{\text{eff}} = 1/2$ and $j_{\text{eff}} = 3/2$ bands, and the preservation of a $4$-in/$4$-out ground state across the whole accessed magnetic region of the phase diagram. ZF (zero-field) $\mu^{+}$SR measurements confirm indeed that the local magnetic configuration is only weakly perturbed by pressure in the $P \lesssim 24$ kbar range.

We measured $M$ (dc magnetization) for a polycrystalline Eu$_{2}$Ir$_{2}$O$_{7}$ sample as a function of $T$ (temperature) for $2$ K $\leq T \leq 150$ K and under hydrostatic pressures $P \lesssim 42$ kbar \cite{SM}. Measurements were performed at a fixed value of the external magnetic field $H = 1$ kOe always by warming the sample after a FC (field-cooling) protocol. Representative results for the magnetic susceptibility $\chi_{m} = M_{m}/H$ in molar formula units at the different investigated $P$ values are shown in Fig.~\ref{GraGaskets}. We define the transition temperature $T_{N}$ to the LRO (long-range ordered) phase from the kink-like anomaly in $M(T)$ (see the double-linear fitting procedure graphically shown in the insets of Fig.~\ref{GraGaskets}). We also define the amount of FC susceptibility associated with the magnetic phase upon investigation as $\Delta\chi_{m} = \chi_{m}(25 \text{ K}) - \chi_{m}(T_{N})$ (see the left panel of Fig.~\ref{GraGaskets}). The choice of $T = 25$ K is due to residual extrinsic background contributions from the pressure cell preventing us from fully exploiting measurements at lower $T$ values. However, $T = 25$ K is safely in the saturation regime for $M$ at ambient $P$ \cite{SM}.

We distinguish between two qualitatively different regimes for Eu$_{2}$Ir$_{2}$O$_{7}$ under $P$ (see Fig.~\ref{GraGaskets}). For $P \lesssim 20$ kbar, both $T_{N}$ and $\Delta\chi_{m}$ increase with increasing $P$, while for $P \gtrsim 25$ kbar both quantities decrease. The situation is summarized later on in Fig.~\ref{GraSumm}, where the actual $P$ dependence of both $T_{N}$ and $\Delta\chi_{m}$ is reported. $T_{N}$ exhibits a smooth and continuous dependence on $P$ clearly passing through a maximum at $P \sim 20 - 25$ kbar. While the crossover between the two regimes in $\Delta\chi_{m}$ is marked by a discontinuity whose origin could be accounted for by an instrumental artefact \cite{FootNote}, the observation of a qualitatively different behaviour of $\Delta\chi_{m}$ for $P \lesssim 20$ kbar and $P \gtrsim 25$ kbar is unambiguous.

\begin{figure}[b!]
	\vspace{5.4cm} \includegraphics{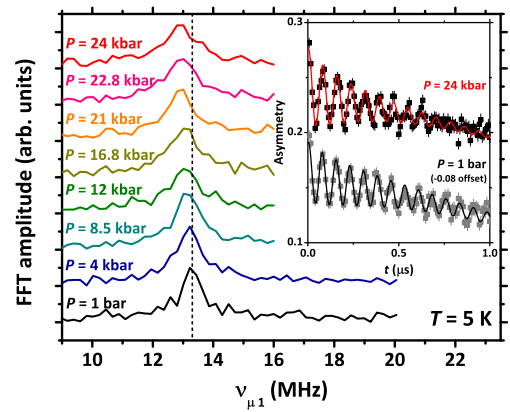}
	\caption{\label{GraSummFFTgpd}(Color online) Main panel: FFT of the $\mu^{+}$ time ZF-depolarizations (representative data are shown in the inset together with their best-fitting curves -- see Ref.~\onlinecite{SM}) measured on the GPD spectrometer at fixed temperature in the saturation regime. The vertical dashed line is a guide-to-the-eye indicating the central frequency at ambient $P$.}
\end{figure}
In order to investigate the effect of $P$ on the magnetic moments' arrangement at the microscopic level, we performed ZF-$\mu^{+}$SR experiments for $P \lesssim 24$ kbar (i.~e., the current maximum value achievable by the employed setup). We show representative raw data for the ZF $\mu^{+}$ spin depolarization in the time domain in the inset of Fig.~\ref{GraSummFFTgpd}. Continuous lines are best-fitting curves according to a standard treatment of ZF-$\mu^{+}$SR data for magnetic materials \cite{SM}. The well-defined spontaneous magnetic field at the muon site $B_{\mu 1}$ is indicative of a LRO magnetic phase \cite{Zha11,Dis12}, resulting in the coherent oscillations shown in the inset of Fig.~\ref{GraSummFFTgpd}. The characteristic frequency $\nu_{\mu 1}$ is directly proportional to the order parameter of the magnetic transition and, in the current case, to the magnetic moment of Ir ions \cite{SM}. From this, we can confirm that no drastic qualitative changes are induced in the Ir magnetic moment by $P$ up to $\sim 24$ kbar.

\begin{figure}[t!]
	\vspace{5.42cm} \includegraphics{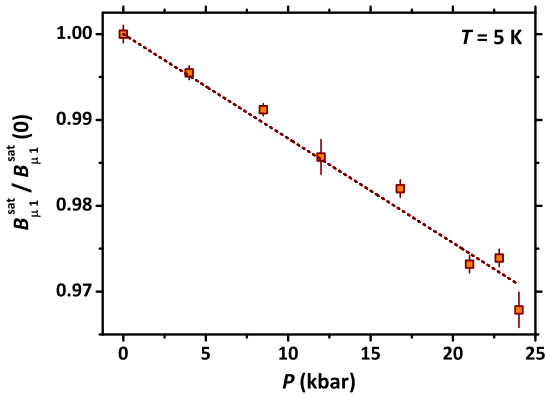}
	\caption{\label{GraSummGPD}(Color online) Local magnetic field at the muon site [normalized to its value $B_{\mu 1}^{\text{sat}} (0) = 977.4$ G at ambient pressure] reported as a function of $P$ after the fitting procedure to experimental ZF time-depolarization curves (see the inset of Fig.~\ref{GraSummFFTgpd}). The dashed line is a linear best-fitting curve to experimental data.}
\end{figure}
A closer examination of data \cite{PraTBP} reveals other features, better visualized by FFT (fast Fourier transform) of the signal in the main panel of Fig.~\ref{GraSummFFTgpd}. A gradual increase of the transversal relaxation, i.~e., the oscillations damping, is observed upon increasing $P$. Accordingly, this is reflected in a progressive broadening of the FFT spectra. We relate this effect to the inevitable progressive increase in $P$ inhomogeneity with increasing $P$ rather than to any intrinsic effect. At the same time, $B_{\mu 1}^{\text{sat}}$ (i.~e., the local field at the muon site in saturation) is continuously shifted to lower values upon increasing $P$. For the aim of clarity, we summarize these latter data in Fig.~\ref{GraSummGPD} clearly showing a linear dependence on $P$. The smoothness of the $B_{\mu 1}^{\text{sat}}$ vs. $P$ trend denotes the absence of any dramatic magnetic or structural effect induced by $P$. Remarkably, we also notice that the relative variation is tiny ($\Delta B_{\mu 1}^{\text{sat}}/B_{\mu 1}^{\text{sat}}(0) \sim 3$ \% at the maximum $P$) and this aspect is not consistent at all with any appreciable change in the Ir$^{4+}$ moment or a change from the $4$-in/$4$-out magnetic configuration. In particular, the observed behaviour could be well explained by a gradual modification in the lattice parameters gradually influencing the muon implantation site and the probed dipolar magnetic field in turn \cite{Pra13a,Bon15}. This argument should be substantiated by {\it ab-initio} calculations of the muon implantation site and by a detailed measurement of the structural parameters of this current sample under comparable $P$ conditions. Unfortunately, the current setup at the GPD facility does not allow to reach higher $P$ values which would allow us to compare data with our results of dc magnetization over the whole $P$ range.

\begin{figure}[b!]
	\vspace{6.1cm} \includegraphics{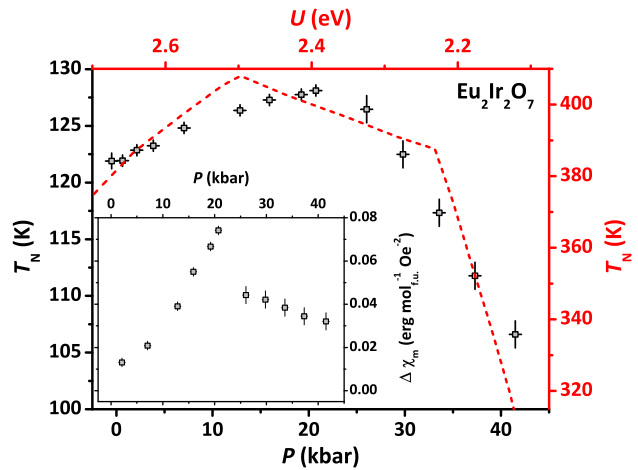}
	\caption{\label{GraSumm}(Color online) Main panel: $P$ dependence of the critical transition temperature (empty squares, bottom and left scales). The red dashed line (top and right scales) is obtained by digitizing data of the calculated EPD discussed in Ref.~\onlinecite{Shi15}. The right scale has been rescaled so that the scale zero coincides with that of the left scale. Inset: $P$ dependence of the FC susceptibility $\Delta\chi_{m}$ proper of the magnetic phase.}
\end{figure}
Remarkably, the non-monotonic behaviour of $T_{N}$ is highly reminiscent of what was recently calculated by means of relativistic LDA-DMFT for the analogue material Y$_{2}$Ir$_{2}$O$_{7}$ \cite{Shi15}. In the current case of Eu$_{2}$Ir$_{2}$O$_{7}$, we exploit the presence of Eu$^{3+}$ at the $R$ site (i.e., a non-magnetic ion such as Y$^{3+}$) allowing us to neglect any $f$--$d$ interaction \cite{Che12} which, although interesting on its own, could in principle hide the intrinsic magnetic properties of the Ir sublattice. We report here data from the calculated EPD \cite{Shi15} superimposed to our experimental results (see the dashed red line in Fig.~\ref{GraSumm}). While the absolute values for the calculated $T_{N}$ are clearly overestimating the experimental result (an issue associated by Shinaoka \textit{et al.} with the neglecting ``of spatial fluctuations in the DMFT approximation'' -- see Ref.~\onlinecite{Shi15}), these can be rescaled by an empirical factor $\sim 3.15$ under the constraint that the scale zero matches the experimental one. Results clearly display a striking agreement with our measurements under the assumption that increasing $P$ is effectively reducing $U/W$ (e.~g., by widening the electronic bandwidth by purely geometrical arguments).

Several interesting observations follow from the results presented above. First of all, the substantial agreement of the experimental and calculated EPDs allows us to rule out that the non-monotonic trend observed for $T_{N}$ is arising from, e.~g., $P$-induced structural phase transitions at $P \sim 25$ kbar but rather points towards an intrinsic underlying electronic mechanism. At the same time, by considering the investigation of transport properties reported in Ref.~\onlinecite{Taf12}, we notice a possible departure of $T_{\text{MI}}$ and $T_{N}$ under $P$ which would solve the current controversy between Mott- vs. Slater-character of the transition in favour of the former. In fact, Tafti \textit{et al.} report that $T_{\text{MI}}$ is mostly unaffected by $P$ up to $\sim 120$ kbar \cite{Taf12}, contrasting with our current reports on $T_{N}$. However, it would be interesting to extend magnetic measurements to higher $P$ values in this respect. Moreover, we stress that the two current datasets are associated with two different samples, making direct comparisons difficult.

The behaviour of $\Delta\chi_{m}$ vs. $P$ needs to be commented further as well. We argue that the initial ($P \lesssim 20$ kbar) fast rise of $\Delta\chi_{m}$ could be due to two different phenomena. On the one hand, $P$ may be inducing different configurations for magnetic domains whose DWs (domain walls) would carry non-trivial exotic properties together with a net magnetization, according to recent theoretical proposals \cite{Yam14}. However, our finding of a LRO phase with non-zero local magnetic field over a bulk fraction of the sample (see below) fails to verify the prediction of a vanishing uniform magnetization in the bulk \cite{Yam14}. Another possible scenario is a gradual modification of the crystalline environment of Ir under $P$ slightly departing from the ideal cubic symmetry. This would imply an increasing degree of canting for the Ir magnetic moments, possibly resulting in an increased magnetic susceptibility in the presence of an external magnetic field. The subsequent decrease of $\Delta\chi_{m}$ with further increasing $P$ can be again understood from Ref.~\onlinecite{Shi15} as an actual decrease of the Ir magnetic moment, whose reduction develops in the same region of the fast suppression of $T_{N}$.

It is worth emphasizing that our experimental validation of the trend calculated in Ref.~\onlinecite{Shi15} suggest a direct verification of other important conclusions of that theoretical study. In the first instance, our data support a departure from the single-band $j_{\text{eff}} = 1/2$ scenario originally proposed for other iridium-based oxides such as, e.~g., Sr$_{2}$IrO$_{4}$ \cite{Kim08} in favour of a hybridized state between the $j_{\text{eff}} = 1/2$ and $j_{\text{eff}} = 3/2$ manifolds. At the same time, independently from the breakdown of the $j_{\text{eff}} = 1/2$ model, we confirm that the $4$-in/$4$-out ground state is preserved across the whole accessed magnetic region of the EPD. Moreover, we also confirmed previous observations by Ishikawa \textit{et al.} \cite{Ish12} about the absence of any thermal hysteresis across $T_{N}$ (not shown), characterizing the phase transition as second-order. This is also in good agreement with the results in Ref.~\onlinecite{Shi15} where the first-order character of the transition is reported only very close to the full disruption of the antiferromagnetic ground state.

Summarizing, in this paper we focussed on the magnetic properties of Eu$_{2}$Ir$_{2}$O$_{7}$ under $P$ reporting on an unprecedented non-monotonic dependence of $T_{N}$ on $P$ and showing that the local magnetic configuration is only weakly perturbed by applied pressure $P \lesssim 24$ kbar. Our results verify the main findings of a recent theoretical study of pyrochlore iridates and, accordingly, they support both a departure from the $j_{\text{eff}} = 1/2$ model conventionally accepted for several iridium-based oxides and the preservation of a $4$-in/$4$-out ground state across the whole accessed magnetic region of the electronic phase diagram.

We thank M. Vojta, J. van den Brink, L. Balents, and Y. Ran for useful discussions and A. Amato, R. Khasanov and H. Luetkens for technical support during the $\mu^{+}$SR measurements. G. Prando acknowledges support by the Humboldt Research Fellowship for Postdoctoral researchers. G. Prando and B. B\"uchner acknowledge support by the Sonderforschungsbereich (SFB) 1143 project granted by the Deutsche Forschungsgemeinschaft (DFG). S. D. Wilson and M. J. Graf acknowledge support by NSF Grant No. DMR-1337567. R. Dally and S. D. Wilson acknowledge support by NSF CAREER Grant No. DMR-1056625. The experimental $\mu^{+}$SR work was performed at the Swiss Muon Source (S$\mu$S) at the Paul Scherrer Institut, Switzerland.



\newpage

\phantom{a}

\newpage


\widetext

\section{\Large{SUPPLEMENTARY MATERIAL}}

\vspace*{0.5cm}

\section{Sample synthesis}

\begin{figure}[htbp]
	\vspace{5.4cm} \includegraphics{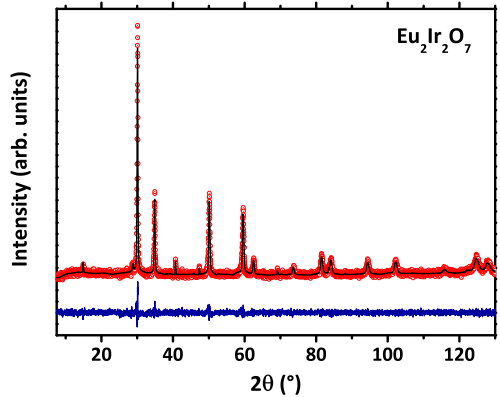}
	\caption{\label{GraXRD}(Color online) Powder XRD diffractogram (experimental data are presented as red circles) together with the relative best-fits after a Rietveld analysis (black line). The blue line shows the residuals after subtraction of data and fit.}
\end{figure}

Stoichiometric amounts of Eu$_{2}$O$_{3}$ (Alfa $99.99$ \%) and IrO$_{2}$ (Alfa $99.99$ \%) were mixed and sintered at $800$ {}$^{\circ}$C for $15$ hours, $1000$ {}$^{\circ}$C for $100$ hours, and $1100$ {}$^{\circ}$C for $100$ hours. The powder was mixed and pelletized with an isostatic cold press in between each sintering step. Sample purity was tested via laboratory XRD (x-ray diffraction) measurements in a Brucker D$2$ Phaser system. Other than a small impurity fraction of $\sim 1$ \% Ir, powder diffraction data (see Fig.~\ref{GraXRD}) revealed a phase pure specimen of Eu$_{2}$Ir$_{2}$O$_{7}$.

\section{Experimental details}

\subsection{dc magnetometry}

\subsubsection{Experimental setup and pressure quantification}

\begin{figure}[t!]
	\vspace{5.4cm} \includegraphics{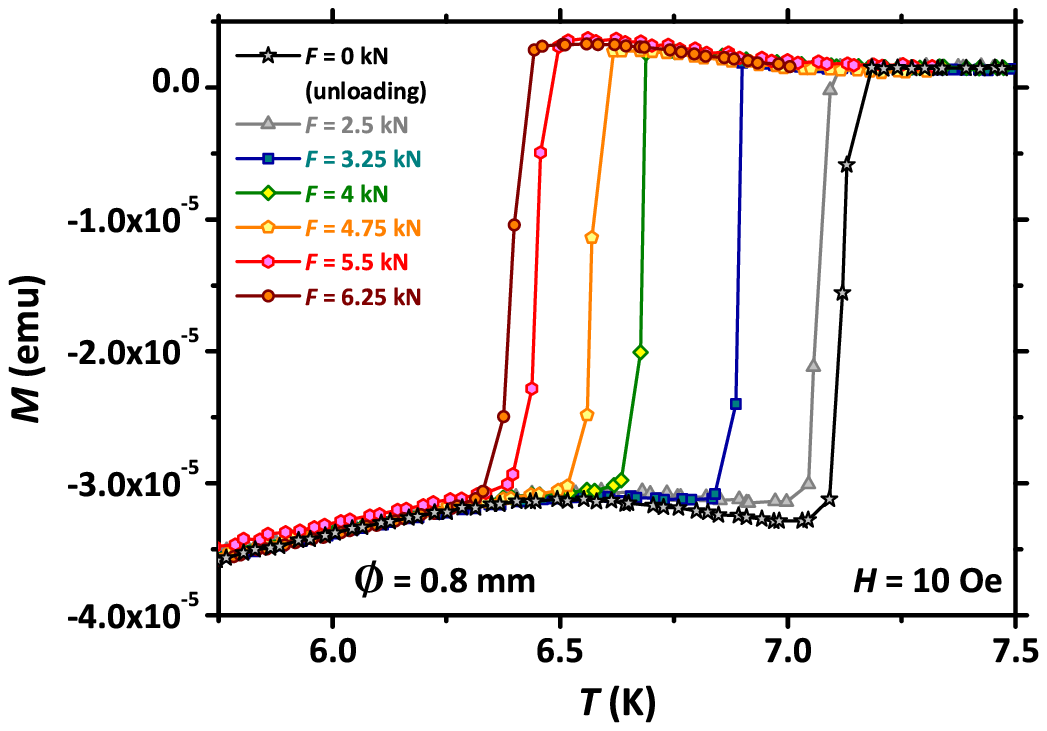} \includegraphics{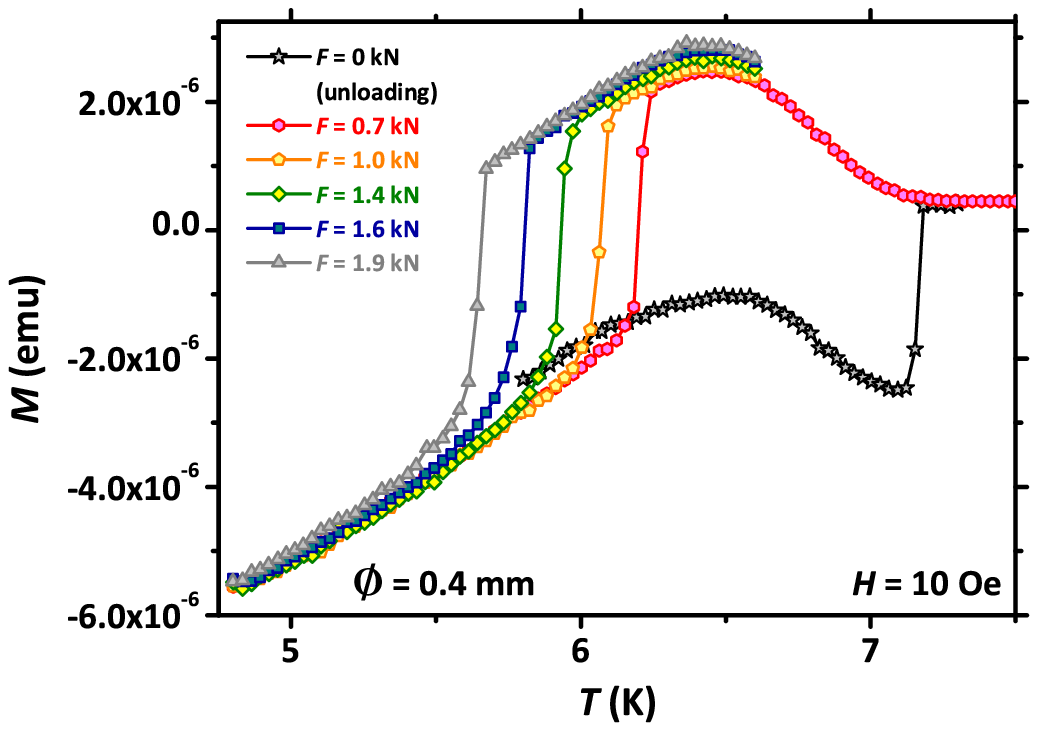}
	\caption{\label{GraLead}(Color online) Superconducting transitions of the Pb manometer for the cases of sample chamber diameter $\varnothing = 0.8$ mm (left) and $\varnothing = 0.4$ mm (right). The different curves correspond to different applied forces (see legend). The sudden jump in magnetization is ascribed to the diamagnetic contribution from the superconducting phase and its overall width is safely below $0.1$ K for all the applied forces, denoting a good pressure homogeneity within the whole accessed experimental $P$ range. Actual $P$ values were quantified from Eq.~\eqref{EqPbSCTransition}. The other source of $T$-dependence for $M$ is the background signal from the PC, different for the two employed setups (see also later).}
\end{figure}

We performed measurements of dc magnetization $M$ by means of a commercial MPMS (magnetic property measurement system) magnetometer by Quantum Design based on a SQUID (superconducting quantum interference device) for temperature values $2$ K $\leq T \leq 150$ K. We employed a home-made CuBe PC (pressure cell) in order to apply external pressure $P$. Here, two opposing cone-shaped ceramic anvils compress a CuBe gasket with a cylindrical hole used as sample chamber \cite{Sch15,Ali09}. The uniaxial force $F$ is applied at ambient temperature and it is converted into hydrostatic pressure in the sample chamber by a homogeneous mixture of ethanol ($1$ part) and methanol ($4$ parts). This choice for the transmitting medium is among the best ones in order to keep the degree of non-hydrostaticity as low as possible within the accessed $P$ range \cite{Pie73}. We employed two different disk-shaped CuBe gaskets with sample space diameter $\varnothing = 0.8$ mm and $\varnothing = 0.4$ mm currently allowing us to reach maximum pressure values $\sim 21$ kbar and $\sim 42$ kbar, respectively. In the case of the former setup, we performed measurements with Daphne oil $7373$ as transmitting medium as well, allowing us to achieve slightly higher maximum $P$ values ($\sim 23$ kbar). In all the cases mentioned above, we checked the $P$ value and its homogeneity at low temperature by measuring the FC (field-cooling) diamagnetic response associated with the superconducting transition of a small Pb manometer inserted in the sample space. We performed these measurements upon the application of a small magnetic field $H = 10$ Oe. Transition widths $\lesssim 0.1$ K denote good $P$ homogeneity at all the investigated $P$ values (see Fig.~\ref{GraLead}). Accordingly, the $P$ value at low $T$ was estimated from the expression \cite{Cla78}
\begin{equation}\label{EqPbSCTransition}
P = \frac{7.17 - T_{c}}{0.0361} \left[\frac{\textrm{K}}{\textrm{K} \times \textrm{kbar}^{-1}}\right],
\end{equation}
$T_{c}$ being the measured mid-point of the diamagnetic transition.

\subsubsection{Subtraction of the background signal from the PC}

\begin{figure}[htbp]
	\vspace{11.3cm} \includegraphics{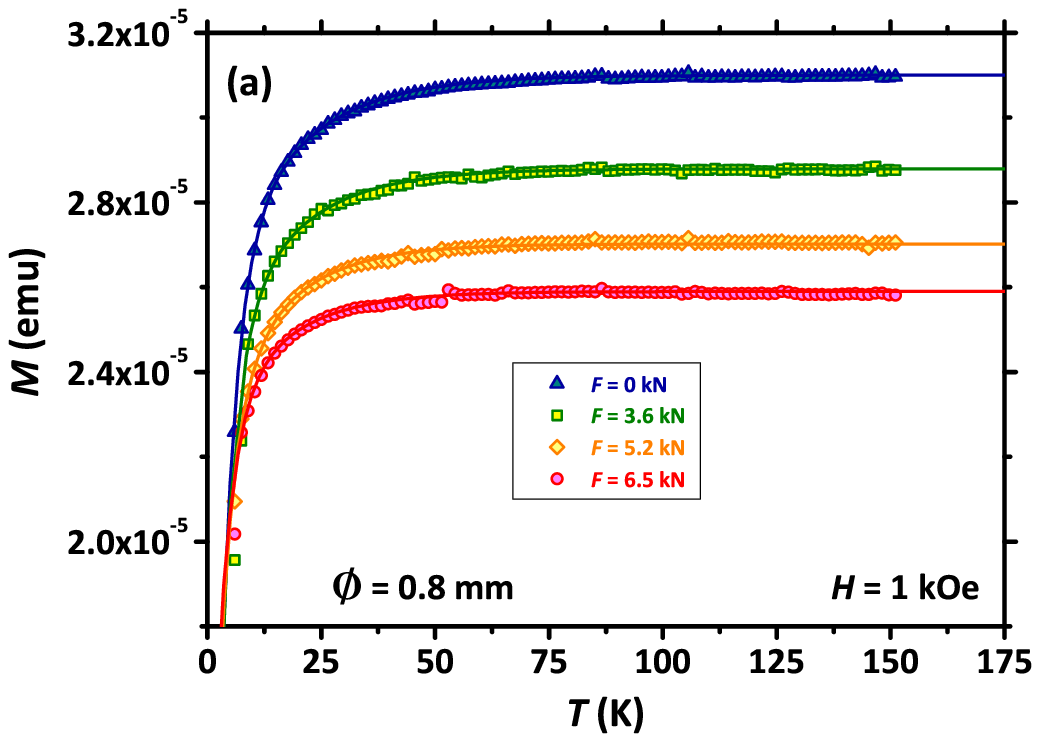} \includegraphics{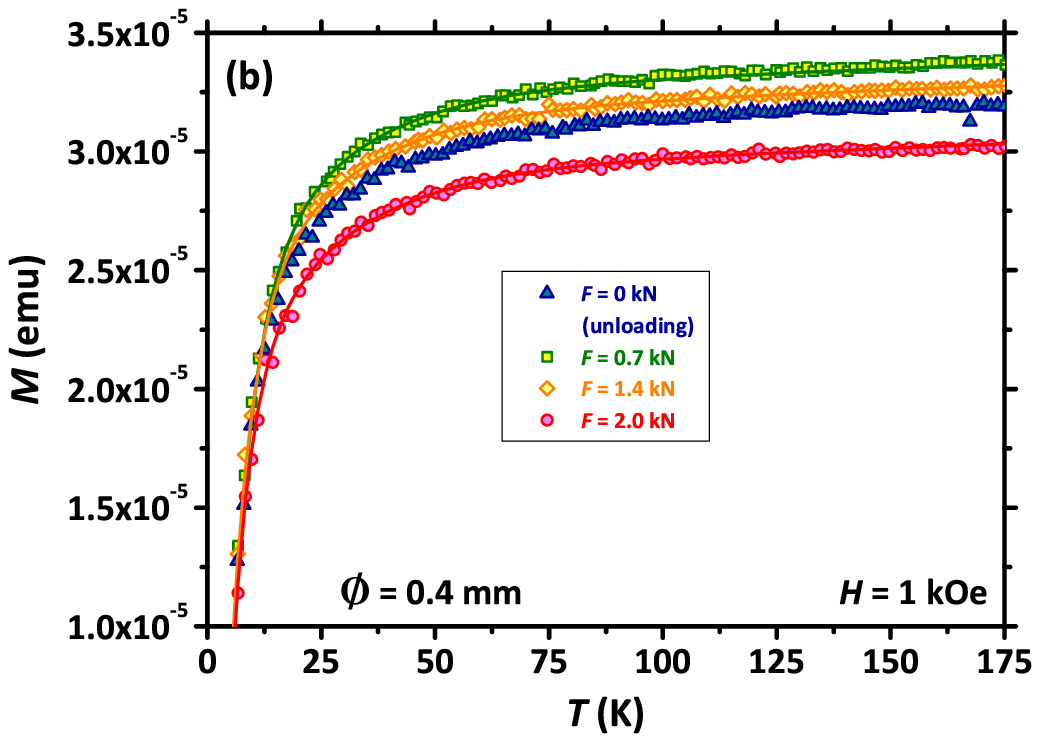}  \includegraphics{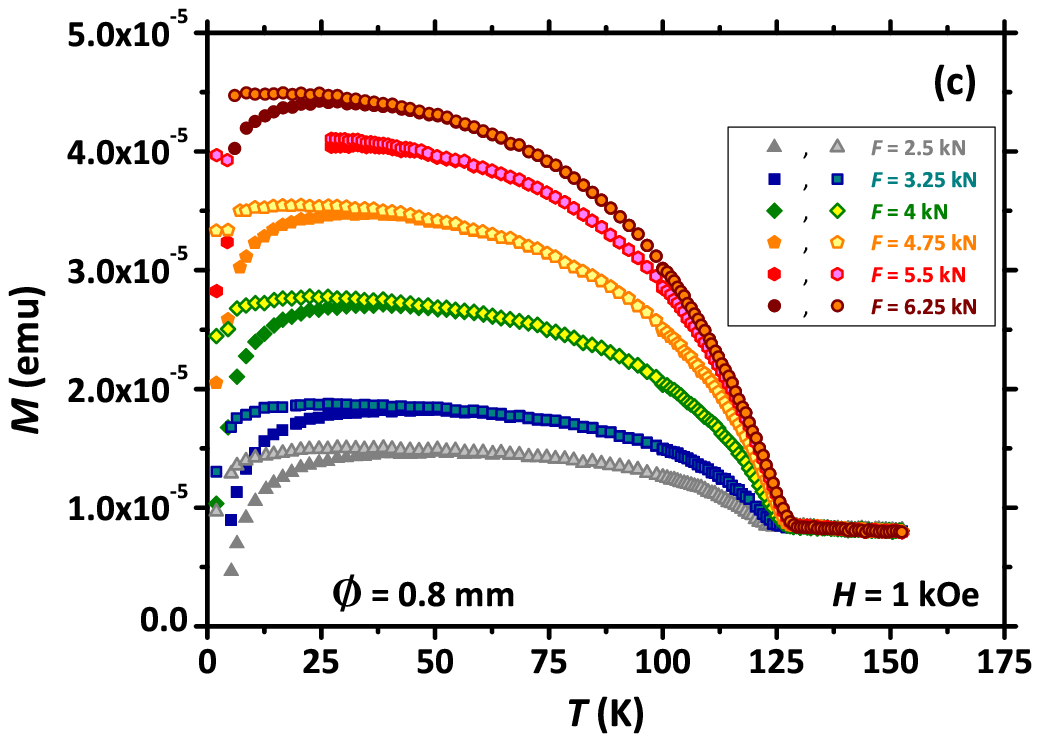} \includegraphics{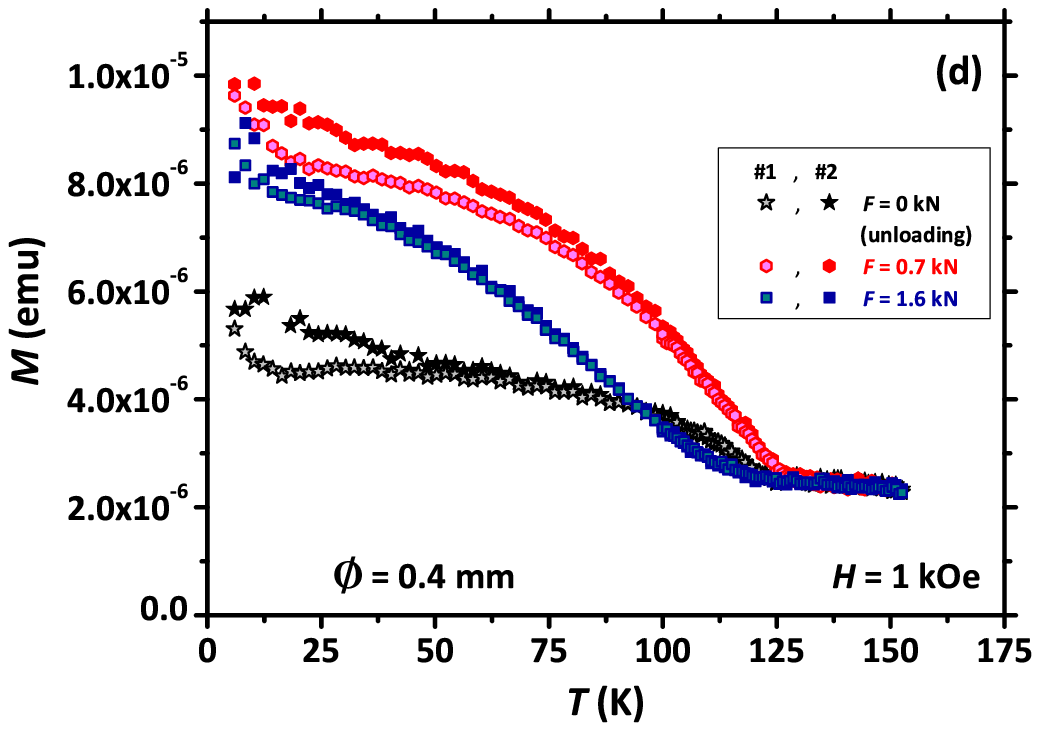}
	\caption{\label{GraSubtraction}(Color online) Representative results of the background subtraction procedure for the magnetization data for both employed setups. The procedure and the meaning of the different symbols are explained in the text below.}
\end{figure}

Measurements of Eu$_{2}$Ir$_{2}$O$_{7}$ were performed at $H = 1$ kOe in FC conditions, most of them upon gradually increasing $P$. The full reversibility of the observed $P$ dependence was checked and verified while unloading the PC, both for the Pb transition and for the actual magnetic properties of the sample. The design of the currently used PC is such that a dramatic reduction of the background signal is achieved which enables measurements of small and weakly magnetic samples. Still, we performed a detailed characterization of the empty PC magnetization within the same experimental conditions for the aim of a quantitative and reliable disentanglement of the sample signal from the overall magnetic response. Representative steps of this procedure are shown in Fig.~\ref{GraSubtraction} and explained below in detail for both employed setups.
\begin{itemize}
	\item The magnetic behaviour of the empty pressure cell is presented in Fig.~\ref{GraSubtraction}(a) and Fig.~\ref{GraSubtraction}(b) for the specific cases of the $\varnothing = 0.8$ mm and $\varnothing = 0.4$ mm gaskets, respectively. Measurements have been performed upon the application of different $F$ in order to check for the effects of gasket distortion on the overall magnetic response. Data have been approximated by an empirical function composed of one linear plus two exponential terms [see the continuous lines in Fig.~\ref{GraSubtraction}(a) and Fig.~\ref{GraSubtraction}(b)] and the extracted characteristic parameters have been used to properly interpolate for the required $F$ values within the successive stages.
	\item In the case of the $\varnothing = 0.8$ mm gasket and within the accessed experimental $T$ range, the absolute variation of the signal from the sample is much stronger than that of the cell background. This can be observed by comparing Fig.~\ref{GraSubtraction}(c) with Fig.~\ref{GraSubtraction}(a). The subtraction of the linear term of the background contribution from the raw data (not shown) results in the full points reported in Fig.~\ref{GraSubtraction}(c). Here, the empty points are the results of subtracting the two exponential terms as well and, after proper normalization by the sample mass, these are presented in Fig.~\ref{GraGaskets}. The background contribution from the PC cannot be fully subtracted at low $T$ values, as observed in Fig.~\ref{GraSubtraction}(c). The actual reason is related to the employment of different gaskets for the background calibrations and the real measurements.
	\item In the case of the $\varnothing = 0.4$ mm gasket the sample amount is smaller and, within the accessed experimental $T$ range, the absolute variation of the signal from the sample is smaller than that of the cell background. This can be observed by comparing Fig.~\ref{GraSubtraction}(d) with Fig.~\ref{GraSubtraction}(b). This fact makes the subtraction procedure more delicate and, for this reason, two different approaches are followed.
	\begin{itemize}
		\item The procedure \#1 is analogous to what is described above for the case of $\varnothing = 0.8$ mm gasket and it gives the results denoted by empty symbols in Fig.~\ref{GraSubtraction}(d). However, small discrepancies between the signal from different gaskets (see above) may be critical.
		\item Within procedure \#2, ZFC (zero-field cooling) data are subtracted from the FC data with the advantage of carefully ``cleaning'' measurements from the cell contribution. This is expected to better account for the $T$ dependence of the background if compared to what performed in procedure \#1 as, in the current case, the signal \textit{from the very same gasket} is being accounted for. Here we exploit the fact that the background term is not exhibiting any irreversible behaviour, i.~e., differences between ZFC and FC branches. However, procedure \#2 will also give a null result above the magnetic transition, where no ZFC/FC opening is observed. A linear contribution is then added to the result in order to match data from procedure \#1 above the magnetic transition. We notice that this procedure, while not rigorous, can be applied to Eu$_{2}$Ir$_{2}$O$_{7}$ after considering its peculiar overall shape of the ZFC curve (see later on, in the right panel of Fig.~\ref{GraMagnCharacterization}). In particular, it is observed that the ZFC branch can be approximately mimicked by a linear contribution over the whole accessed $T$ range (from $2$ K up to $300$ K). The overall result of the subtraction procedure \#2 is shown in Fig.~\ref{GraSubtraction}(d) by full symbols.
	\end{itemize}
	As is shown, the discrepancies between the two methods are essentially irrelevant at relatively high temperatures, while the low-temperature side is still dominated by the differences between the gaskets used for calibration and for actual measurements. In the main paper, results from both the procedures have been used (see Fig.~\ref{GraGaskets} -- the same meaning for symbols is used).
\end{itemize}

\subsection{Muon spin spectroscopy ($\bm{\mu^{+}}$SR)}

We performed $\mu^{+}$SR measurements \cite{Blu99,Yao11} at the S$\mu$S muon source (Paul Scherrer Institut, Switzerland) on the GPD spectrometer ($\mu$E1 beamline) for $5$ K $\leq T \leq 200$ K, in ZF conditions. We applied external pressures at ambient $T$ by means of a double-wall piston-cylinder PC made of MP$35$N alloy. The transmitting medium Daphne oil $7373$ assured nearly-hydrostatic $P$ conditions in the whole experimental range. The $P$ value and its homogeneity was checked at low $T$. In particular, the diamagnetic response associated with the superconducting transition of a small In wire in the sample space was measured by means of ac susceptibility (not shown).

We performed the analysis of our ZF-$\mu^{+}$SR data similarly to what is discussed in, e.~g., Refs.~\cite{Pra13a} and \cite{Pra15b}. The general expression
\begin{equation}\label{EqGeneralFittingZFPCandSample}
A_{T}(t) = A_{0} \left[a_{\textrm{PC}} \; e^{-\frac{\sigma_{\textrm{PC}}^{2} t^{2}}{2} - \lambda_{\textrm{PC}} t} + \left(1 - a_{\textrm{PC}}\right) G_{T}^{\textrm{s}}(t)\right]
\end{equation}
was employed for all the investigated $T$ values as the best-fitting function for the $t$ (time) dependence of the measured asymmetry function $A_{T}$. The maximum value for $A_{T}$ (measured at $t = 0$) is a spectrometer-dependent parameter denoted as $A_{0}$ and it corresponds to full spin polarization ($\sim 100$ \%) for the implanted $\mu^{+}$ (positive muons). It can be shown \cite{Blu99,Yao11} that the quantity $P_{T} \equiv A_{T}/A_{0}$ defines the spin (de)polarization for the implanted $\mu^{+}$. $a_{\textrm{PC}} \simeq 0.5$ -- $0.65$ mainly accounts for the fraction of incoming $\mu^{+}$ stopping in the PC [when performing ambient $P$ calibration experiments on the low-background GPS spectrometer ($\pi$M3 beamline), $a_{\textrm{PC}} \simeq 0$ arises from the $\mu^{+}$ implanted into the sample holder, into the cryostat walls, etc]. Within the PC, $P_{T}$ is affected by the nuclear magnetism of the MP$35$N alloy. This leads to a damping governed by the Gaussian and Lorentzian parameters $\sigma_{\textrm{PC}}$ and $\lambda_{\textrm{PC}}$, whose $T$ dependence was independently estimated in a dedicated set of measurements on the empty PC. In the current case of Eu$_{2}$Ir$_{2}$O$_{7}$, the function
\begin{eqnarray}\label{EqGeneralFittingZFSample}
G_{T}^{\textrm{s}}(t) & = &
\left[1 - V_{\textrm{m}}(T)\right] e^{-\frac{\sigma_{\textrm{N}}^{2} t^{2}}{2}} +{}\nonumber\\ & + & \left[a^{\parallel}(T) D^{\parallel}(t) + \sum_{i=1}^{2} a_{i}^{\perp}(T) F_{i}(t) D_{i}^{\perp}(t)\right]
\end{eqnarray}
was employed in order to describe the spin (de)polarization arising from the remaining fraction $\left(1 - a_{\textrm{PC}}\right)$ of $\mu^{+}$ (i.~e., $\mu^{+}$ implanted into the sample). $V_{\textrm{m}}(T)$ is the fraction of $\mu^{+}$ probing static local magnetic fields and, due to the random implantation of $\mu^{+}$, it is equivalent to the magnetic volume fraction of Eu$_{2}$Ir$_{2}$O$_{7}$. For $V_{\textrm{m}}(T) = 0$ (high temperatures), only nuclear magnetic moments can cause a damping of the signal. Typically this damping is of Gaussian nature with a characteristic rate $\sigma_{\textrm{N}} \sim 0.1 \; \mu$s$^{-1}$. Below the critical transition temperature to the magnetic phase, the superscript $\parallel$ ($\perp$) refers to $\mu^{+}$ probing a local static magnetic field in a parallel (perpendicular) direction with respect to the initial $\mu^{+}$ spin polarization. Accordingly, one has $\left[a^{\parallel}(T) + \sum_{i} a_{i}^{\perp}(T)\right] = V_{\textrm{m}}(T)$ for the so-called ``longitudinal'' ($a^{\parallel}$ or $a^{L}$) and ``transversal'' ($a_{i}^{\perp}$ or $a_{i}^{Tr}$) fractions. Here, $i = 1,2$ denotes two different inequivalent crystallographic sites of implantation for $\mu^{+}$, as reflected in two different signals at very low-$t$ values (in agreement with the previous observations in Ref.~\cite{Zha11}). Our experimental results suggest a different statistical population of these two sites ($a_{1}^{\perp}/a_{2}^{\perp} \sim 4$).

The longitudinal components typically probe dynamical spin-lattice-like relaxation processes resulting in slow exponentially-decaying functions $D_{i}^{\parallel}(t) = e^{-\lambda_{i}^{L}t}$. Due to the typically low values measured for $\lambda_{i}^{L}$ ($\lesssim 0.1$ $\mu$s$^{-1}$) in comparison with the overall experimental $t$-window ($\sim 5$ $\mu$s), the two different $i$ components cannot be resolved in the longitudinal fraction and only one averaged $D^{\parallel}(t) = e^{-\lambda^{L}t}$ is reported in Eq.~\eqref{EqGeneralFittingZFSample}, accordingly. On the other hand, the $\perp$ component yields valuable information about the static magnetic properties of the investigated phase. A coherent precession of $\mu^{+}$ around the local field $B_{\mu}$ generated by a LRO (long-range ordered) magnetic phase (static within the $\mu^{+}$ mean lifetime $\sim 2.2$ $\mu$s) can be observed in the transversal amplitude indeed. This is typically described by oscillating functions $F_{i}(t)$, with the common choice $F_{i}(t) = \cos\left(\gamma B_{\mu i} t + \phi_{i}\right)$, $\gamma = 2 \pi \times 135.54$ $\mu$s$^{-1}$/T being the gyromagnetic ratio for $\mu^{+}$ and $\phi_{i}$ phase terms. At the same time, the damping functions $D_{i}^{\perp}(t) = e^{-\lambda_{i}^{Tr}t}$ reflect a static distribution of local magnetic fields at the site of the $\mu^{+}$. In the current case of Eu$_{2}$Ir$_{2}$O$_{7}$, the oscillating signal from the $i = 1$ site is clearly visible and allows us to quantify both $B_{\mu 1}$ (proportional to the order parameter of the magnetic phase) and its distribution $\sqrt{\langle B_{\mu 1}^{2} \rangle} \sim \lambda_{1}^{Tr}/\gamma$. On the other hand, we find that $\lambda_{2}^{Tr}$ is so high to cause an overdamping of oscillations for this site and, accordingly, we set $F_{2}(t) = 1$. Otherwise stated, we find $B_{\mu 1} \gg \lambda_{1}^{Tr}/\gamma$ and $B_{\mu 2} \lesssim \lambda_{2}^{Tr}/\gamma$.

Summarizing, the fitting function we employed for ZF-$\mu^{+}$SR measurements of Eu$_{2}$Ir$_{2}$O$_{7}$ can be re-written as
\begin{eqnarray}\label{EqGeneralFittingZFSampleRewr}
G_{T}^{\textrm{s}}(t) & = &
\left[1 - V_{\textrm{m}}(T)\right] e^{-\frac{\sigma_{\textrm{N}}^{2} t^{2}}{2}} +{}\nonumber\\ & + & \left[a^{\parallel}(T) e^{-\lambda^{L}t} + a_{1}^{\perp}(T) \cos\left(\gamma B_{\mu 1} t + \phi_{1}\right) e^{-\lambda_{1}^{Tr}t} + a_{2}^{\perp}(T) e^{-\lambda_{2}^{Tr}t}\right].
\end{eqnarray}
The fitting results are in good agreement with our experimental data at all $T$ and $P$ values (statistical $\chi^{2} \simeq 1 - 1.1$).

\section{Sample characterization at ambient pressure}

\subsection{Transport properties}

\begin{figure}[htbp]
	\vspace{5.6cm} \includegraphics{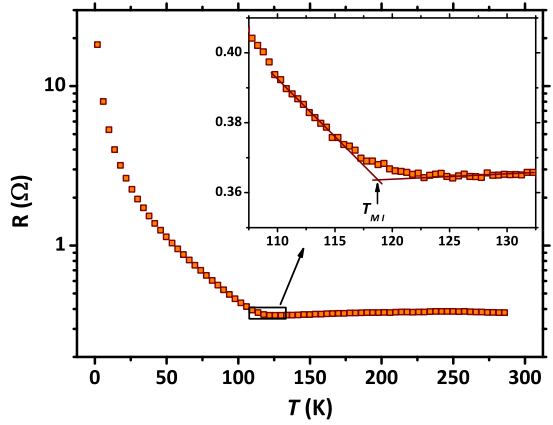}
	\caption{\label{GraResistivity} (Color online) Resistance vs. $T$ for the current Eu$_{2}$Ir$_{2}$O$_{7}$ polycrystalline sample at ambient $P$ (the inset shows an enlargement in the transition region. The arrow shows the estimated $T_{\text{MI}}$ value).}
\end{figure}

Fig.~\ref{GraResistivity} shows the $T$ dependence of the resistance $R$ of our currently investigated Eu$_{2}$Ir$_{2}$O$_{7}$ sample. Measurements were performed at ambient $P$ and in ZF conditions with a $4$-terminal ac setup (the dependence on the current amplitude was checked). A clear anomaly is detected at $T_{\text{MI}} = \left(118.8 \pm 0.3\right)$ K, corresponding to the metal-to-insulator transition (the inset shows an enlargement of the transition region and the graphical procedure to estimate $T_{\text{MI}}$). $R$ strongly increases with decreasing $T$ below $T_{\text{MI}}$ while a conventional $dR/ d T > 0$ metallic state is observed above the transition.

\subsection{Magnetic properties}

\begin{figure}[b!]
	\vspace{5.4cm} \includegraphics{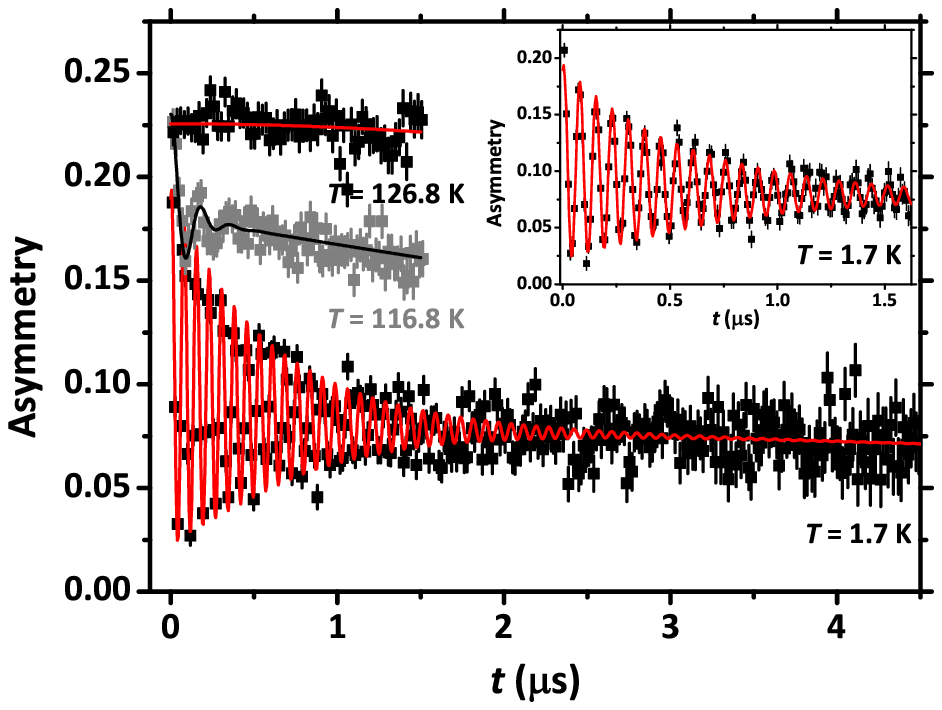} \includegraphics{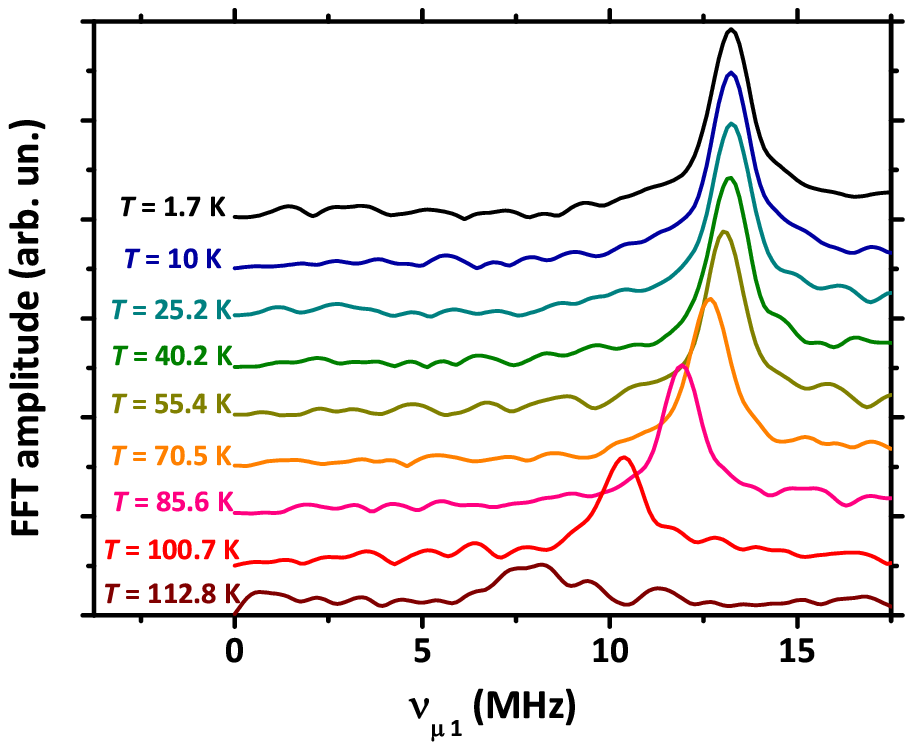}
	\caption{\label{GraMusrRawData}(Color online) Representative ZF-$\mu^{+}$SR depolarization curves of Eu$_{2}$Ir$_{2}$O$_{7}$ as measured on GPS in the time domain (left) and after a FFT procedure (right). In the left panel, continuous lines are best-fits according to Eq.~\eqref{EqGeneralFittingZFPCandSample} and Eq.~\eqref{EqGeneralFittingZFSampleRewr}, where $a_{\textrm{PC}} \simeq 0$. The inset shows an enlargement of the $T = 1.7$ K curve at short times. In the right panel, we have set $\nu_{\mu 1} = \gamma B_{\mu 1} / 2\pi$.}
\end{figure}
Representative ZF-$\mu^{+}$SR $t$-depolarization curves for Eu$_{2}$Ir$_{2}$O$_{7}$ measured on the low-background spectrometer GPS are reported in the left panel of Fig.~\ref{GraMusrRawData}. The sudden development of a static internal magnetic field $B_{\mu 1}$ is clearly observed below $T_{N} \sim 120$ K, being reflected in the appearance of long-lived coherent oscillations in $A_{T}(t)$. As can be clearly observed both in the time and frequency domain (after a FFT, Fast Fourier Transform, of the signal), the condition $B_{\mu 1} \gg \lambda_{1}^{Tr}/\gamma$ discussed above is well verified (see the right panel of Fig.~\ref{GraMusrRawData}). The central frequency at low $T$ is $\gamma B_{\mu 1} / 2\pi \equiv \nu_{\mu 1} = 13.24(5)$ MHz, in good agreement with the value $\nu_{\mu 1} = 13.32(3)$ MHz from previous reports \cite{Zha11}. The transversal component from the second site ($a_{2}^{\perp}$) is damped with the characteristic rate $\lambda_{2}^{Tr} \sim 4$ $\mu$s$^{-1}$ and, as such, it only involves very short times of the depolarization (see also Fig.~1 in Ref.~\cite{Zha11}). Concerning the phase value, we systematically observe negative values $\left|\phi_{1}\right| \sim 20$°.

\begin{figure}[t!]
	\vspace{5.4cm} \includegraphics{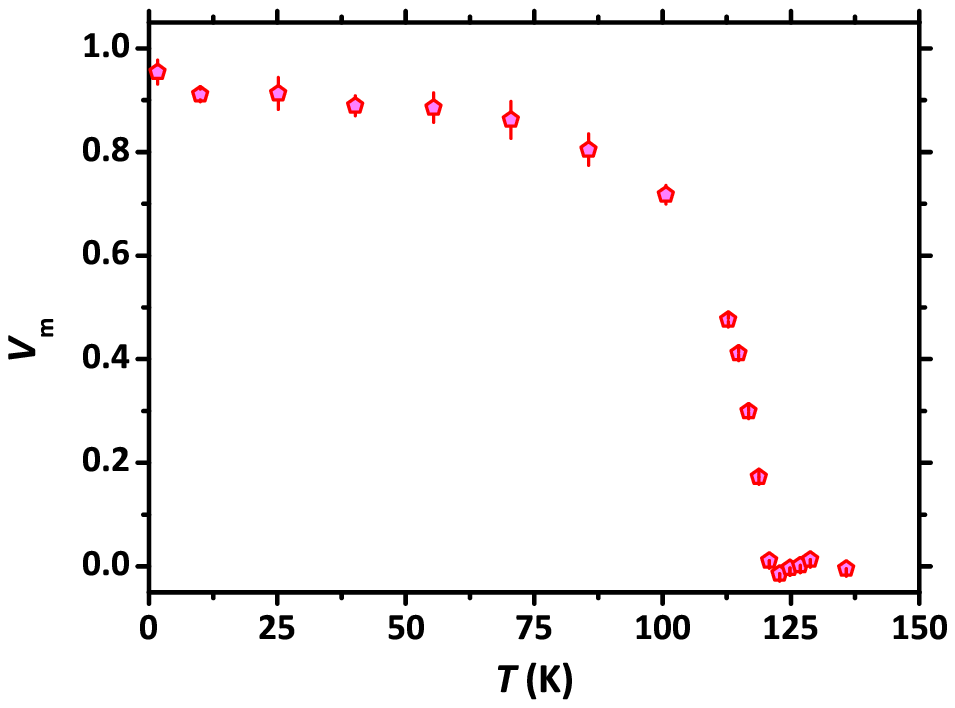} \includegraphics{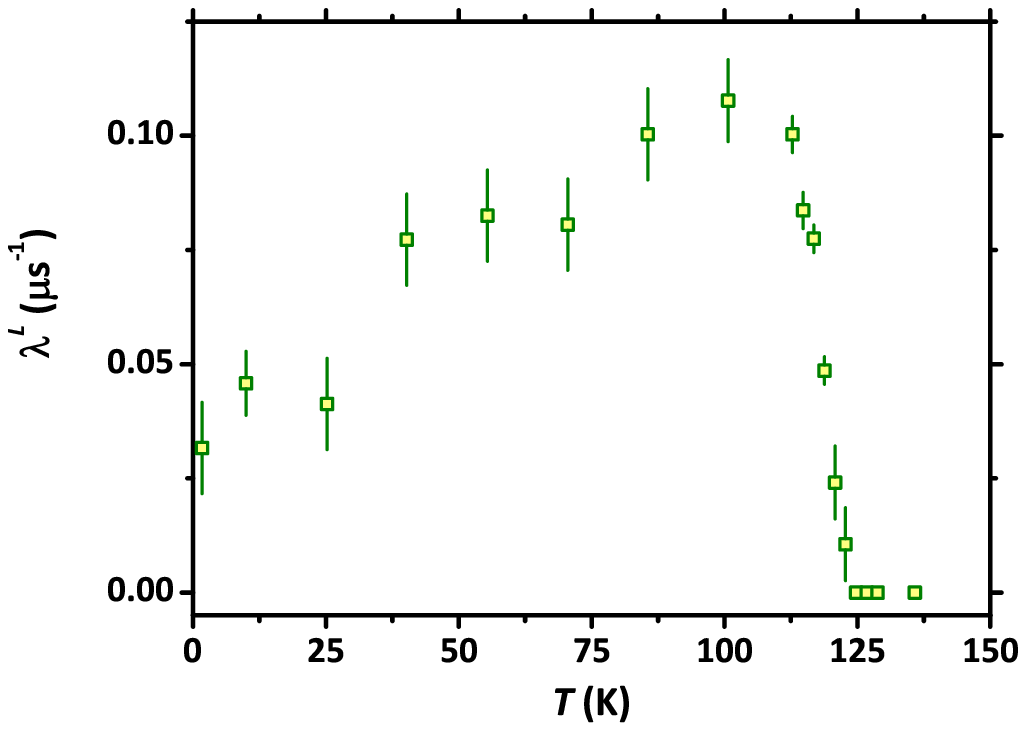}
	\caption{\label{GraMusrSummarizing}(Color online) $T$ dependence for the magnetic volume fraction $V_{\textrm{m}}$ (left) and for the longitudinal dynamical relaxation rate $\lambda^{L}$ (right). Both the datasets have been obtained after fitting Eq.~\eqref{EqGeneralFittingZFPCandSample} and Eq.~\eqref{EqGeneralFittingZFSampleRewr} to the experimental depolarization curves in Fig.~\ref{GraMusrRawData} in the time-domain.}
\end{figure}
\begin{figure}[t!]
	\vspace{5.4cm} \includegraphics{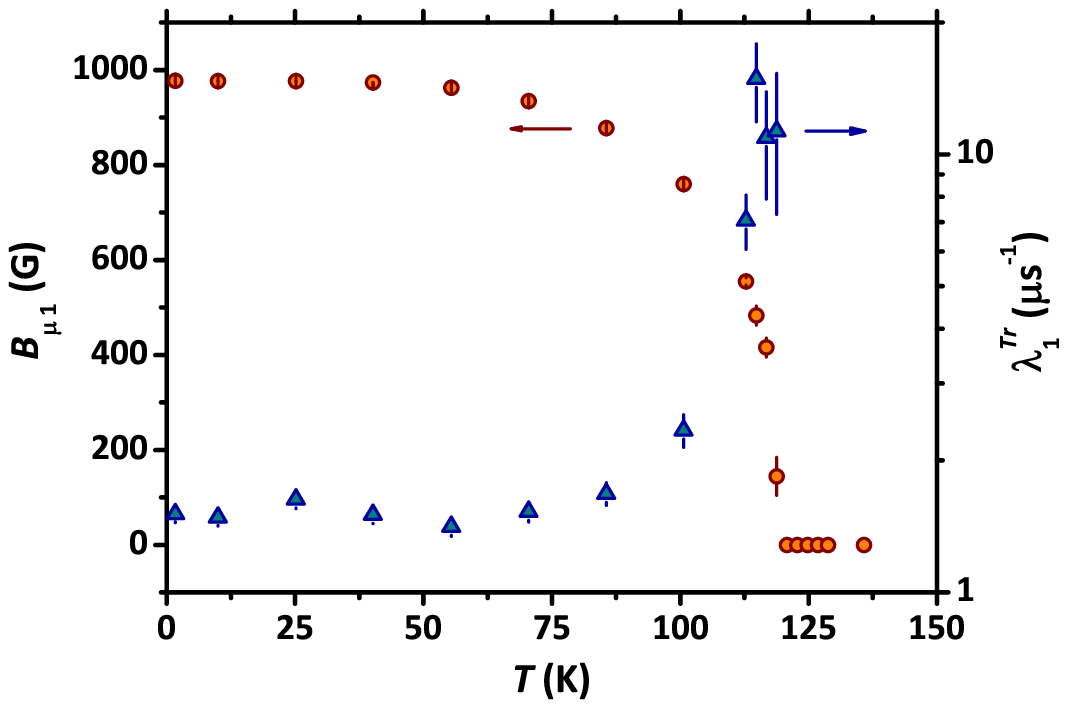} \includegraphics{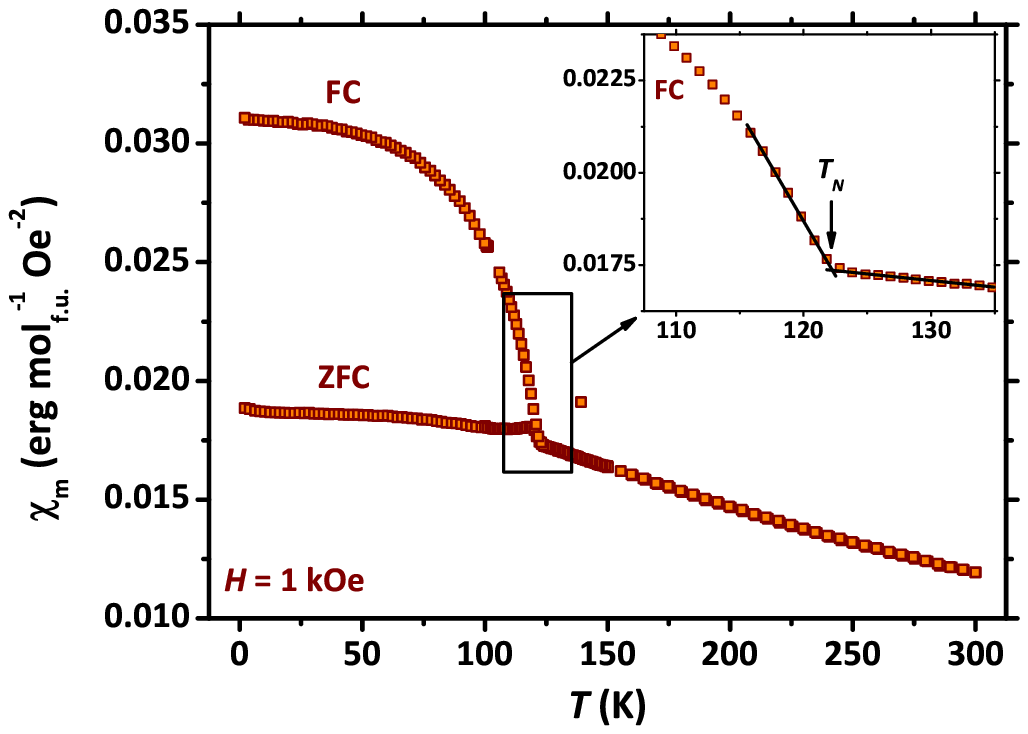}
	\caption{\label{GraMagnCharacterization}(Color online) Left panel: $T$ dependence for the internal magnetic field $B_{\mu 1}$ (left scale) and the transversal damping parameter $\lambda_{1}^{Tr}$ (right scale). Both the datasets have been obtained after fitting Eq.~\eqref{EqGeneralFittingZFPCandSample} and Eq.~\eqref{EqGeneralFittingZFSampleRewr} to the experimental depolarization curves in Fig.~\ref{GraMusrRawData} in the time-domain. Right panel: molar susceptibility $\chi_{m} = M_{m} / H$ as a function of $T$ at $H = 1$ kOe both in ZFC and FC conditions. Inset: enlargement around the sharp anomaly observed in the FC curve. The double-linear graphical fitting procedure allows one to define the value of the critical temperature to the magnetic phase, $T_{N}$ (see arrow).}
\end{figure}

Fig.~\ref{GraMusrSummarizing} and Fig.~\ref{GraMagnCharacterization} show the $T$ dependence of the parameters derived from fitting Eq.~\eqref{EqGeneralFittingZFPCandSample} and Eq.~\eqref{EqGeneralFittingZFSampleRewr} to the time-domain depolarization curves in the left panel of Fig.~\ref{GraMusrRawData}. The left panel of Fig.~\ref{GraMusrSummarizing} displays the magnetic volume fraction $V_{\textrm{m}}(T)$. Here, we observe a sudden development of a bulk magnetic fraction below $T_{N} \sim 120$ K, saturating at $V_{\textrm{m}}(T) \sim 0.9 - 0.95$, in good agreement with what is also reported in Ref.~\cite{Zha11}. The remaining non-magnetic fraction can be well accounted by the background amplitude $a_{\textrm{PC}}$ plus a tiny segregated non-magnetic phase in the sample. As commented in the main text of the paper, it should be remarked that the actual shape of $V_{\textrm{m}}(T)$ is quite unusual. It was commented in Ref.~\cite{Zha11} that this shape for $V_{\textrm{m}}(T)$ may arise from a statistical distribution of different values for the transition temperature. However, this scenario seems quite unrealistic in view of the good static properties of the internal field (discussed below). Moreover, in the presence of an inhomogeneous distribution of transition temperatures, one would expect a symmetrical shape for $V_{\textrm{m}}(T)$ around the mid-point of the overall transition with the typical functional form \cite{Pra13b}
\begin{equation}\label{EqMagneVolERFC}
V_{\textrm{m}}(T) = \frac{1}{2} \; \textrm{erfc}\left[\frac{T - T_{N}} {\sqrt{2}\Delta}\right],
\end{equation}
the complementary error function $\textrm{erfc}(x)$ being
\begin{equation}
\textrm{erfc}(x) = \frac{2}{\sqrt{\pi}} \int_{x}^{+\infty}
e^{-t^{2}} dt.
\end{equation}
At the same time, we report in the right panel of Fig.~\ref{GraMusrSummarizing} the $T$ dependence of the longitudinal dynamical relaxation rate $\lambda^{L}$. Remarkably, and different from what has been reported previously \cite{Zha11}, we can clearly detect a well-defined critical peak when approaching the magnetic transition from above, followed by a slow decrease approaching $\lambda^{L} = 0.03$ $\mu$s$^{-1}$ with further decreasing $T$. While the sharp increase at $T \sim 120$ K is the expected behaviour for canonical magnetic phase transitions, the slow decrease below $T_{N}$ is more unconventional and points towards the persistence of anomalous spin dynamics within the magnetic phase.

The static local quantities probed by $\mu^{+}$ are reported in Fig.~\ref{GraMagnCharacterization} (left panel). The $T$ dependence of $B_{\mu 1}$ is in good agreement with what has been reported previously \cite{Zha11}. At the same time, the transversal damping parameter $\lambda_{1}^{Tr}$ saturates at low $T$ at $\lambda_{1}^{Tr} \simeq 1.5$ $\mu$s$^{-1}$, a value almost $3$ times smaller than what has been reported in Ref.~\cite{Zha11}. This can be already seen in the inset of Fig.~\ref{GraMusrRawData} (left panel), where clear oscillations are still well detected at $t = 1.5$ $\mu$s. Together with the observation of a dynamical critical peak (see Fig.~\ref{GraMusrSummarizing}), these indications generally suggest better magnetic properties for the currently investigated sample if compared to previous results in Ref.~\cite{Zha11}.

The dependence of the molar magnetic dc susceptibility $\chi_{m} = M_{m} / H$ on temperature is displayed in the right panel of Fig.~\ref{GraMagnCharacterization}, showing a behaviour in excellent agreement with previous reports \cite{Tai01,Ish12}. The absence of sizeable upturns of $\chi_{m}$ for $T \leq 10$ K denotes negligible contributions from possible spurious magnetic phases in our currently investigated sample \cite{Tai01,Ish12}. We observe a well-defined opening of ZFC and FC curves at $T_{N} = \left(122.0 \pm 0.2\right)$ K, which is in good agreement with what is observed from ZF-$\mu^{+}$SR results. This value has been estimated from the double-linear procedure graphically shown in the inset of Fig.~\ref{GraMagnCharacterization} (right panel). Throughout the main text of this paper, this is the adopted definition for $T_{N}$.


\begin{thebibliography}{99}
\bibitem{FootNote} In particular, the observed behaviour may be associated with the two experimental setups employed for dc magnetization measurements, as discussed in Ref.~\onlinecite{SM}. This issue is currently still not completely understood and will be the subject of further experimental investigation.
\end{thebibliography}
\end{document}